\documentclass[%
 aip,
 amsmath,amssymb,
preprint,%
]{revtex4-1}

\usepackage{graphicx}% Include figure files
\usepackage{dcolumn}% Align table columns on decimal point
\usepackage{bm}% bold math
\usepackage[utf8]{inputenc}
\usepackage[T1]{fontenc}
\usepackage{mathptmx}

\begin{document}

%\preprint{AIP/123-QED}

\title{Scalable creation of silicon-vacancy color centers in diamond by ion implantation through a 1-$\mu$m pinhole}
% Force line breaks with \\

\author{L.~Hunold}
\email{lukas.hunold@student.uni-siegen.de}
\affiliation{ 
Laboratory of Nano-Optics and C$\mu$, University of Siegen, Siegen, Germany
}%
\author{S.~Lagomarsino}%
\affiliation{ 
Laboratory of Nano-Optics and C$\mu$, University of Siegen, Siegen, Germany
}%
\affiliation{ 
Istituto Nazionale di Fisica Nucleare, Sezione di Firenze,  Sesto Fiorentino, Italy
}%
%\altaffiliation[Also at ]{Istituto Nazionale di Fisica Nucleare, Sezione di Firenze, Sesto Fiorentino, Italy.}%Lines break automatically or can be forced with \\
\author{A.M.~Flatae}
\affiliation{ 
Laboratory of Nano-Optics and C$\mu$, University of Siegen, Siegen, Germany
}%
\author{H.~Kambalathmana}
\affiliation{ 
Laboratory of Nano-Optics and C$\mu$, University of Siegen, Siegen, Germany
}%
\author{F.~Sledz}
\affiliation{ 
Laboratory of Nano-Optics and C$\mu$, University of Siegen, Siegen, Germany
}%
\author{S.~Sciortino}%
\affiliation{ 
Department of Physics and Astronomy, University of Florence, Sesto Fiorentino, Italy
}%
\affiliation{ 
Istituto Nazionale di Fisica Nucleare, Sezione di Firenze,  Sesto Fiorentino, Italy
}%
\author{N.~Gelli}%
\affiliation{ 
Istituto Nazionale di Fisica Nucleare, Sezione di Firenze,  Sesto Fiorentino, Italy
}%
\author{L.~Giuntini}%
\affiliation{ 
Department of Physics and Astronomy, University of Florence, Sesto Fiorentino, Italy
}%
\affiliation{ 
Istituto Nazionale di Fisica Nucleare, Sezione di Firenze, Sesto Fiorentino, Italy
}%
\author{M.~Agio}
\email{mario.agio@uni-siegen.de}
\homepage{https://nano-optics.physik.uni-siegen.de.}
\affiliation{ 
Laboratory of Nano-Optics and C$\mu$, University of Siegen, Siegen, Germany
}%
\affiliation{ 
National Institute of Optics (INO), National Research Council (CNR), Florence, Italy
}%

\date{\today}% It is always \today, today,

\begin{abstract}
The controlled creation of quantum emitters in diamond represents a major research effort in the fabrication of single-photon devices. Here, we present the scalable production of silicon-vacancy (SiV) color centers in single-crystal diamond by ion implantation. The lateral position of the SiV is spatially controlled by a 1-$\mu$m pinhole placed in front of the sample, which can be moved nanometer precise using a piezo stage. The initial implantation position is controlled by monitoring the ion beam position with a camera. Hereby, silicon ions are implanted at the desired spots in an area comparable to the diffraction limit. We discuss the role of ions scattered by the pinhole and the activation yield of the SiV color centers for the creation of single quantum emitters. 
\end{abstract}

\maketitle

%\begin{quotation}
%The ``lead paragraph'' is encapsulated with the \LaTeX\ \verb+quotation+ environment and is formatted as a single paragraph before the first section heading. (The \verb+quotation+ environment reverts to its usual meaning after the first sectioning command.) Note that numbered references are allowed in the lead paragraph.
%The lead paragraph will only be found in an article being prepared for the journal \textit{Chaos}.
%\end{quotation}

%\section{Introduction}

Color centers in diamond are considered a very promising platform for quantum photonics devices due to their optical properties and the possibility to create diamond nanostructures~\cite{pezzagna_creation_2011,schroder_quantum_2016,sipahigil_integrated_2016,aharonovich_diamond_2018,sipahigil_quantum_2019}. Among them, the negatively-charged silicon-vacancy (SiV) center has already achieved a number of important goals, such as bright emission concentrated in the zero-phonon line (ZPL)~\cite{neu_photophysics_2012}, two-photon interference~\cite{sipahigil_indistinguishable_2014}, spin preparation and readout~\cite{rogers_all-optical_2014}, coherent dipole-dipole interaction~\cite{evans_photon-mediated_2018}, electroluminescence~\cite{lohrmann_diamond_2011} and single-photon emission in n-type diamond~\cite{flatae_silicon-vacancy_2020}, temperature sensitivity~\cite{nguyen_all-optical_2018} and photostability even at high temperatures~\cite{lagomarsino_robust_2015}.

Color centers can be obtained either during diamond growth~\cite{neu_silicon_2012} or by ion implantation followed by thermal annealing~\cite{lagomarsino_optical_2018}. Only in the second case their position can be laterally controlled, for instance by focused ion-beam techniques~\cite{tamuraArrayBrightSiliconvacancy2014,schroder_scalable_2017,zhouDirectWritingSingle2018} or by aperture-type AFM tips~\cite{raatz_investigation_2019-1}, which is crucial for device fabrication. Although these techniques have demonstrated very high lateral resolution (down to about 10 nm), so far they have only been investigated at low ion energies (a few keV), which limits the implantation depth to only a few nanometers. It is therefore desirable to study the creation of color centers, in particular the SiV, in a broader range of ion energies to gain flexibility in terms of implantation depths.

Here, we implement a commercial 1-$\mu$m pinhole and a high-precision translation stage into a beamline of a tandem accelerator to implant Si ions in spots comparable to the diffraction limit. This is done in a controlled manner, and in a wide range of energies (0.4-3 MeV) as well as fluences (10$^8$-10$^{14}$ cm$^{-2}$). In this way, we present a flexible and easy implementable approach for a localized implantation process. We further investigate the role of ions scattered by the pinhole to determine the conditions for the creation of single SiV centers. Finally, we demonstrate the scalable generation of single emitters by measuring their antibunching characteristics.

\begin{figure}
\includegraphics[width=0.45\textwidth]{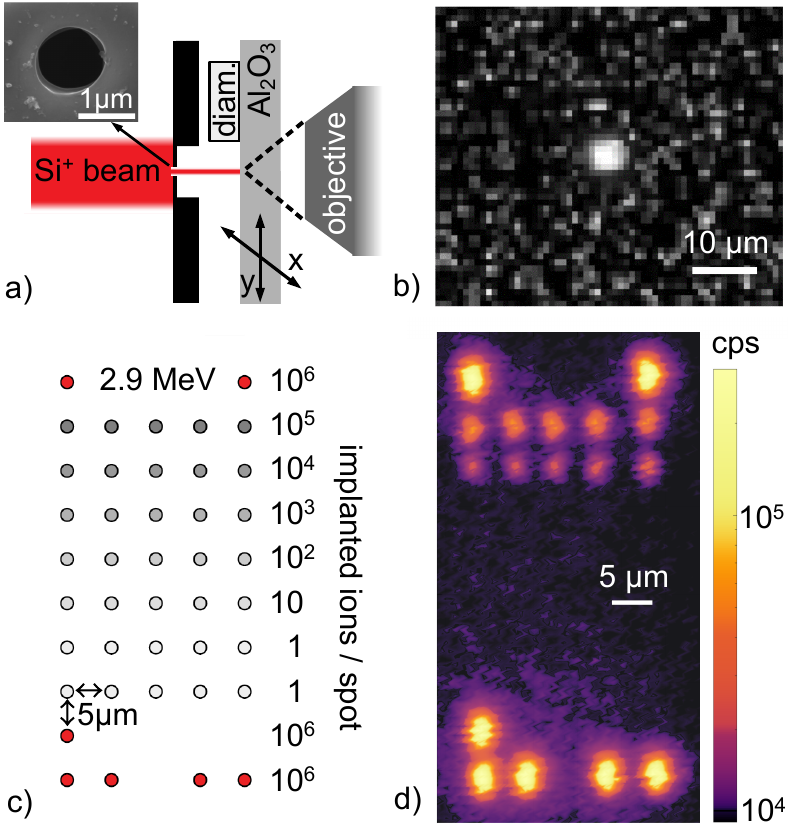}
\caption{\label{fig:fig1} (a) Implantation setup with pinhole, diamond sample on sapphire plate and objective sending the PL created by the beam to a camera. Sample and plate can be moved in $x$, $y$ (perp. to the beam) with a piezo stage. Inset shows an SEM image of the 1 $\mu$m pinhole. (b) Beam image on the camera used to set the initial implantation position. (c) Plan of implantation session A (compare Tab.~\ref{tab:table1} for session labeling). (d) Corresponding PL map acquired by confocal scan.}
\end{figure}

%\section{Ion-implantation}
The implantation facility is described in detail in Ref.~\onlinecite{lagomarsino_center_2018}. In short terms, it consists of a 3 MV tandem accelerator employing a cesium sputter ion source able to form ion beams from solid state samples. The maximum energy depends on the ion state of charge, and amounts to 15 MeV in the case of silicon. Implantation takes place in a vacuum chamber downstream to the accelerator. The facility was previously used for versatile SiV color center creation~\cite{lagomarsino_creation_2021} and is now upgraded with a system for localized implantation as described in the following. Annealing of the centers at $1200~^\circ$C is carried out in an alumina oven surrounded by graphitic heaters, all placed in a stainless steel ultra-high vacuum chamber.

Figure~\ref{fig:fig1}a shows a scheme of the implantation setup. The beam enters the configuration from the left and passes the pinhole to reach a sapphire (Al$_2$O$_3$) plate. Here it produces a photoluminescence (PL) signal, which is monitored by a CMOS camera (ZWO ASI178MM) using a microscope objective. The PL spot is enlarged to about 5 pixels (see Fig.~\ref{fig:fig1}b) in order to clearly discriminate it from hot pixels on the camera. The initial implantation position is then set by moving the sample into the beam until the desired location is reached. The absolute accuracy of this overall process is evaluated to approx. $\pm 5$ $\mu$m and it is limited by the resolution of the camera. The relative positioning of the following implantation spots is then very accurate and limited by the precision of the translation stage (Q-545.240, Physik Instrumente GmbH, 26 mm travel range, 1 nm resolution).

The commercial pinhole is from Thorlabs (P1H) and it has a nominal diameter of 1 $\mu$m, with a tolerance of +0.25/-0.1 $\mu$m and a circularity of $\ge$ 80\%. The scanning electron microscopy (SEM) picture (Fig.~\ref{fig:fig1}a inset) indicates a smooth circular geometry and a diameter slightly larger than 1 $\mu$m, but within the specified tolerance. The foil thickness around it was measured optically to $\mathrm{d}_\mathrm{foil}$ = (27.5 $\pm$ 1.1) $\mu$m, which is important to evaluate a possible effect of a pinhole tilt on the ion throughput. We emphasize here, that for our 1 $\mu$m hole, this issue is not as critical as in the case of nano-apertures~\cite{raatz_investigation_2019-1}. Nonetheless, the relatively thick frame of the pinhole demands a precise perpendicular mounting relative to the incoming ion beam. For the given system the horizontal and vertical tilt was measured to < $0.3^\circ$, suitable for this implantation process.

\begin{table}
\caption{\label{tab:table1}Implantation sessions and corresponding parameters.}
\begin{ruledtabular}
\begin{tabular}{ccccc}
Implantation&Energy&Fluence&Separation&Rows x Columns\\
Session&(MeV)&(cm$^{-2}$)&($\mu$m)&\\ 
\hline 
A&2.9&10$^{8}$-10$^{14}$&5&10 x 5\\
B&0.4&10$^{8}$-10$^{14}$&5&10 x 5\\
C&1&10$^{9}$-10$^{11}$&10&3 x 8 \\
D&0.4&10$^{10}$&10&1 x 5 \\
\end{tabular}
\end{ruledtabular}
\end{table}

Figure~\ref{fig:fig1}c illustrates a typical plan for an implantation matrix on an electronic grade diamond sample (ElementSix, 2$\times$2$\times$0.5 mm$^3$, face orientation 110, roughness < 5 nm). The beam fluence is varied in order to generate spots with 10$^6$ to about 1 implanted ion with an energy of 2.9 MeV, corresponding to an implantation depth of around 1.1 $\mu$m. Matrices with other parameters were created as well, the four most important ones for this work are summarized in Table \ref{tab:table1}. 

After annealing, the PL signal of the formed emitters is acquired using a confocal scanning setup described in detail in  Ref.~\onlinecite{kambalathmana_optical_2021}. Figure~\ref{fig:fig1}d shows an exemplary result, reproducing the implanted matrix except for the spots with 10$^3$ Si ions or less. These are not visible, because their PL signal is overcome by a background noise associated with the diffuse implantation of ions scattered at the pinhole. This issue is discussed further below and was studied in detail to eventually enable the identification of spots down to single emitters.

%\section{Results and discussion}

\begin{figure}
\includegraphics[width=0.45\textwidth]{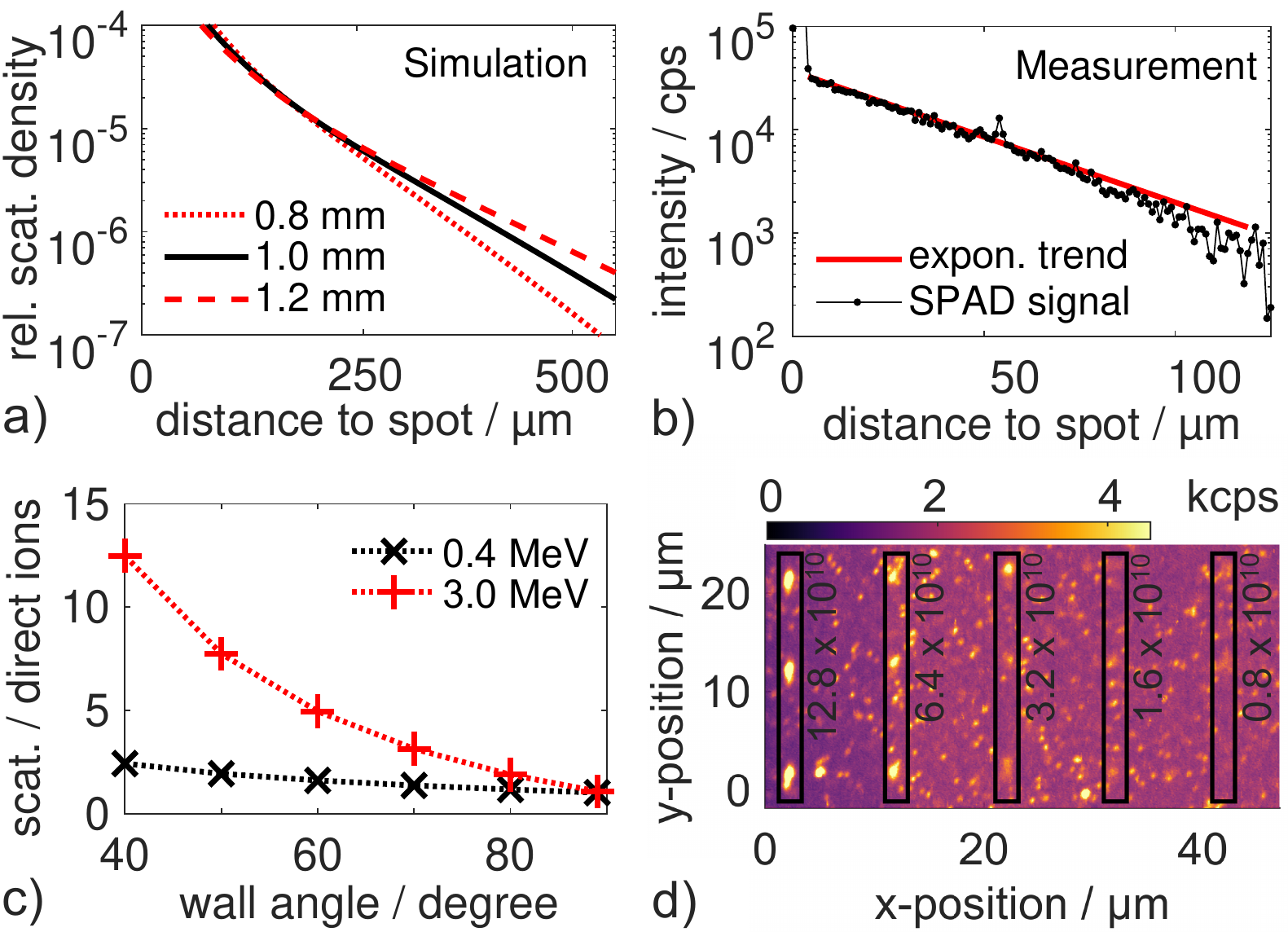}
\caption{\label{fig:fig2} (a) Simulated distance dependencies of scattered ions for the case of 2.9 MeV, 40$^\circ$ pinhole wall angle and three different pinhole-sample distances corresponding to the used value of (1$\pm$0.2) mm. (b) SiV background measured on the sample for one of the strongest spots in session A. (c) Ratio of total scattered to total direct ions for ion energies of 0.4 MeV and 3 MeV as a function of the pinhole wall angle. (d) Confocal scan of the matrix associated with session C, showing scattered emitters around the desired spots. Fluences of the spots in the different columns are given in the figure (units cm$^{-2}$).}
\end{figure}

Ions are scattered towards the sample by the pinhole, when its edges have thin parts allowing ions to be transmitted and distracted from their initial path by collisions within the material. We verify that this effect is relevant for our implantations by comparing the SiV signal around the created spots with simulations of the process carried out with the SRIM tool~\cite{ziegler_srim_2010}. For this, we model the used pinhole with a conical shape and determine the amount of ions passing through its thin parts for a given energy. Then we simulate the resulting trajectories after the pinhole and the corresponding distribution of ions on the sample for a given distance. In principle, larger distances are favorable, since they result in lower densities of scattered ions, but in the experiment not more than 1 mm was used to avoid increased spot sizes due to the beam divergence (measured to about 0.3 mrad). 

We use this value of 1 mm pinhole-sample distance and an energy of 2.9 MeV for the described simulation. This results in the distance dependence of the scattered ion density relative to the ion density in the implanted spot as shown in Fig.~\ref{fig:fig2}a. The experimental counterpart was obtained by measuring the SiV signal while moving away from the implanted spot (Fig.~\ref{fig:fig2}b). A similar exponential decrease of the background resulting from scattered ions and thereby created SiV emitters was found, supporting the model assumptions. It was also found that the measured background drops faster than expected, which might be attributed to uncertainties in the pinhole-sample distance or the simulations at the given energy and for the used pinhole material (steel). 

The absolute strength of the shown background is determined by the number of scattered ions per directly implanted ion and depends strongly on the ion energy and pinhole wall angle (Fig.~\ref{fig:fig2}c). The wall shape can not be adjusted for the given pinhole and also not determined precisely. However, Figure~\ref{fig:fig2}d indicates a small angle corresponding to strong scattering, visible as several bright spots around the implanted matrix points of session C. Based on these results, the energy was chosen to 0.4 MeV for the creation of single emitters (session D) and the stronger spots needed for initial finding of the matrix were located further away (400 $\mu$m distance).

\begin{figure}
\includegraphics[width=0.45\textwidth]{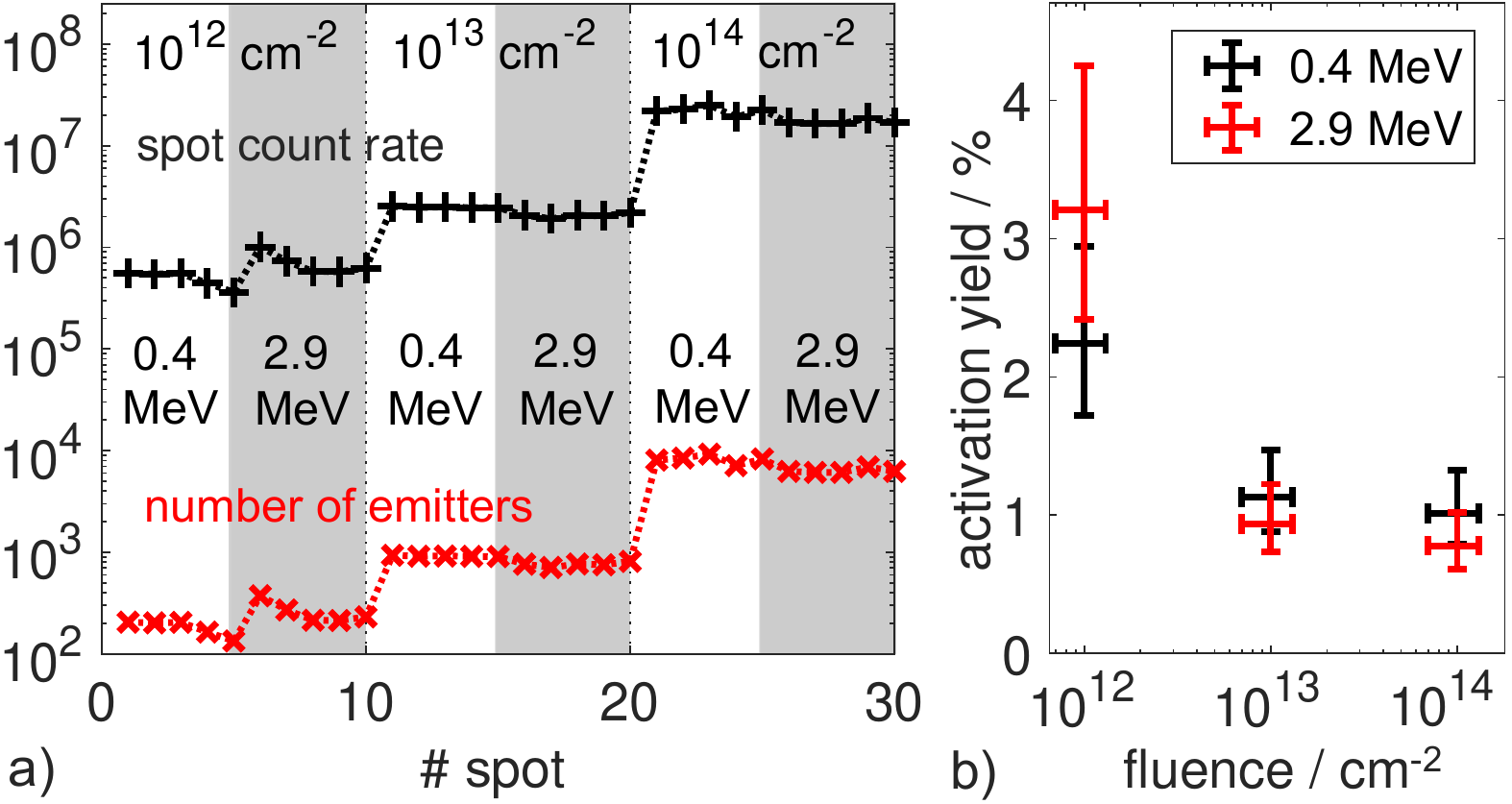}
\caption{\label{fig:fig3} (a) Count rates (black curve) and corresponding emitter numbers (red curve) for highly implanted spots. Ten points were investigated for each fluence from $10^{12}$ cm$^{-2}$ over $10^{13}$ cm$^{-2}$ to $10^{14}$ cm$^{-2}$, as indicated by the values given in the figure. From those, always five were implanted with ion energies of 0.4 MeV (white columns) and five with 2.9 MeV (gray columns). (b) Concluded activation yield of centers depending on energy and dose.}
\end{figure}

The remaining task is to identify the appropriate beam fluence for the creation of a single emitter per spot (on average). For this the spots with a high number of emitters were studied first, which allow to quantify the activation yield of the color center conversion process. By using confocal intensity scans of the differently strong implanted spots, their SiV emission count rate was determined (Fig.~\ref{fig:fig3}a black curve). With the two-dimensional intensity map it is possible to take into account that the laser only excites part of the spot when focused on it (by scaling up the signal with the ratio between spot and laser focus size). 

Next, the number of emitters corresponding to these signals is calculated by using the average count rate of a single emitter as reference. This was estimated by averaging the signal of more than 300 bright spots found around the matrix of session C (compare Fig.~\ref{fig:fig2}d). By measuring the emission lifetime and spectrum of several random sample spots, it was verified that these are SiV color centers created by scattered ions. The relatively large separation (few micrometers) allows the assumption that most of them are single centers. In addition, spots with a size clearly larger than the diffraction limit were excluded. Therefore, the average signal of the remaining spots is seen as a good estimate for the single emitter count rate, which was evaluated to (2700$\pm$300) cps under the given experimental conditions. Here, the used excitation laser power (3.3 mW at the sample surface), the excitation wavelength (656 nm), the NA of the focusing objective (0.95) and the overall collection and detection efficiency of the system (estimated to 0.12 \%) are especially important. 

By maintaining these parameters for the measurements on the strongly implanted spots, the number of emitters per spot (Fig.~\ref{fig:fig3}a red curve) can be calculated. Based on this, the SiV activation yield is concluded for the given fluences (Fig.~\ref{fig:fig3}b). Here the uncertainties are mainly due to instabilities in the beam current (error bars on the fluences) and an imprecise determination of the pinhole throughput (error bars on the activation yield). The values are in agreement with those reported in the literature~\cite{schroder_scalable_2017,lagomarsino_optical_2018}. Beyond that, they provide information about the SiV activation yield for the given energies, which have not been studied in detail before. 

Next, the determined activation yield allows us to predict the number of created emitters also for spots with very low ion densities. If the activation remains constant for fluences below $10^{12}$ cm$^{-2}$, a value of 0.6$\times 10^{10}$ cm$^{-2}$ would provide roughly one activated emitter per spot on average. The assumption of constant activation yield at lower ion doses is supported by previous studies.~\cite{schroder_scalable_2017} To verify this, implantation session C was used. It consist of eight columns of emitter spots, starting with a fluence of 1.28$\times 10^{11}$ cm$^{-2}$ and reducing it by a factor of two for any of the following columns. Thereby transition between clear spots and single emitters is immediately evident (see Fig.~\ref{fig:fig2}d). From the confocal scan, the column with 0.8$\times 10^{10}$ cm$^{-2}$ appears to be the first which reaches single emitters in the spots, supporting the results obtained by extrapolation from the strongly implanted spots.

\begin{figure}
\includegraphics[width=0.45\textwidth]{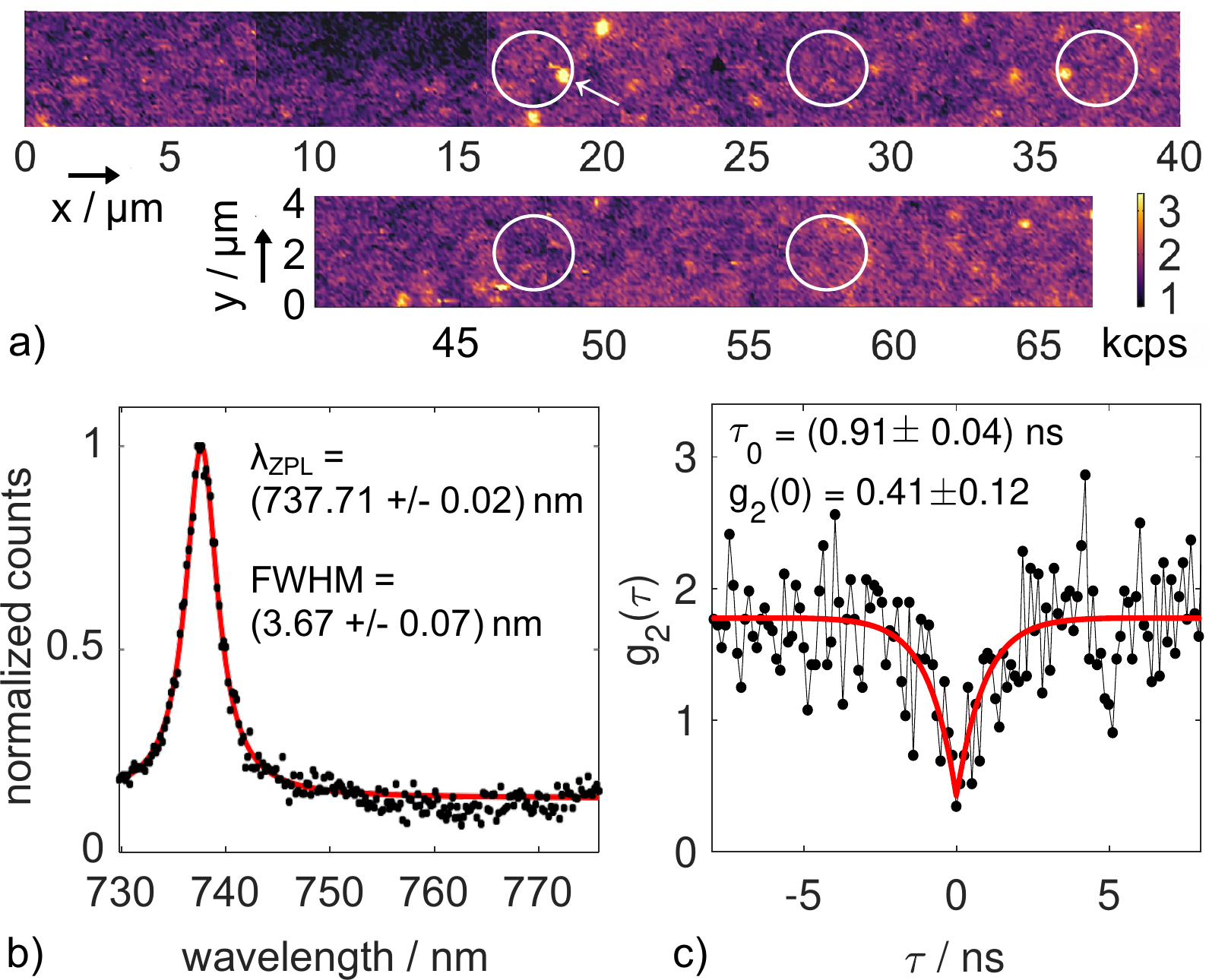}
\caption{\label{fig:fig4} (a) Confocal scan of the region around the matrix associated with session D. The expected positions of the matrix spots are indicated by the white circles. One of the identified emitters is marked with a white arrow. (b) Emission spectrum of the marked spot. The Lorentzian fit shows a clear ZPL with the parameters given in the figure. (c) Corresponding second order correlation measurement verifying the creation of a single SiV emitter by the presence of antibunching (parameters given in the figure).}
\end{figure}

Although some uncertainties are involved in the estimations above, the result allows the creation of spots with single emitters, because the number of activated centers per spot is assumed to follow a Poissonian distribution, as found also e.g., in Ref.~\onlinecite{tamuraArrayBrightSiliconvacancy2014}. Based on this, a mean number of 0.5-2 emitters per spot will statistically result in 25-35\% spots with a single emitter. Therefore, single emitter creation is also possible under the above discussed conditions.

As proof of principle, five spots were created with a fluence of 1.6$\times 10^{10}$ cm$^{-2}$ in the context of session D. The confocal scan reveals almost no background associated with scattered ions (Fig.~\ref{fig:fig4}a), in contrast to the previous approaches. This indicates that the adjustments discussed above (lower ion energy, stronger spots at a larger distance) could solve the problem. A few bright spots were identified, out of which two were located roughly at the expected positions (white circles). We further investigate them by measuring their spectral and temporal emission characteristics. The Lorentzian fit shows a clear ZPL (5-6 times the sample background) around 738 nm, proving the desired creation of SiV color centers (Fig.~\ref{fig:fig4}b). Furthermore, the second order correlation measurements reveal antibunching behavior, verifying single photon emission (Fig.~\ref{fig:fig4}c). Based on this, we conclude the general possibility of using the given concept for the localized creation of single color centers.

%\section{\label{sec:level1}Conclusions}

In conclusion, we discuss the creation of SiV color centers by ion implantation through a 1-$\mu$m pinhole at desired locations in spots with size comparable to the diffraction limit. We identify conditions for obtaining single emitters, including discussions of ion scattering by the pinhole and the activation yield. Hence, we demonstrate the ability to create SiV centers in a scalable manner, in a wide range of depths (energies).
Our approach can be further improved by reducing the pinhole diameter (higher lateral precision) and the use of vertical pinhole walls (less ion scattering). Moreover, the technique is general and can be applied to other ion species like e.g. nitrogen or germanium and other host matrices such as e.g. silicon carbide. 

%\begin{acknowledgments}
The authors acknowledge financial support from the University of Siegen and the German Research Foundation (DFG) (INST 221/118-1 FUGG, 410405168). Experimental support was provided by N. Soltani (optical setup) and M. Hepp (SEM images) in association with the Micro- and Nanoanalytics Facility of the University of Siegen. The authors also acknowledge INFN-CHNet, the network of laboratories of the INFN for cultural heritage, for support and precious contributions in terms of instrumentation and personnel. The authors wish to thank F. Taccetti for experimental assistance and suggestions.

The data that support the findings of this study are available from the corresponding author upon reasonable request.
%\end{acknowledgments}

%\appendix

%\section{Appendixes}

%To start the appendixes, use the \verb+\appendix+ command.

%\bibliography{references}% Produces the bibliography via BibTeX.


\end{document}